\documentclass[%
letter,
twocolumn,
 superscriptaddress,
 amsmath,amssymb,
 aps,
 aip,
 nofootinbib
]{revtex4-2}

\usepackage{graphicx}
\usepackage{dcolumn}
\usepackage{bm}
\usepackage{xcolor}
\usepackage{ulem}
\usepackage{hyperref}
\usepackage{cleveref}
\usepackage{amsmath}
\usepackage{float}
\usepackage{svg}
\usepackage{amssymb}
\usepackage{scalerel}
\usepackage{footmisc}

\begin{document}

\title{Electrical control of single-photon emission in highly-charged individual colloidal quantum dots}

{
\makeatletter
\def\frontmatter@thefootnote{%
 \altaffilletter@sw{\@fnsymbol}{\@fnsymbol}{\csname c@\@mpfn\endcsname}%
}%
\makeatother

\author{Sergii Morozov}
\altaffiliation[Present address:]{ Centre for Nano Optics, University of Southern Denmark, Campusvej 55, Odense M, DK-5230, Denmark}
\affiliation{The Blackett Laboratory, Department of Physics, Imperial College London, London SW7~2BW, United Kingdom}
\author{Evangelina L. Pensa}
\altaffiliation[Present address:]{ Physics of Synthetic Biological Systems, Physics-Department and ZNN, Technische Universit\"at M\"unchen, Garching 85748, Germany
}
\affiliation{The Blackett Laboratory, Department of Physics, Imperial College London, London SW7~2BW, United Kingdom}
\author{Ali Hossain Khan}
\affiliation{Department of Chemistry, Ghent University, Krijgslaan 281-S3, Gent 9000, Belgium}

\author{Anatolii Polovitsyn}
\affiliation{Department of Chemistry, Ghent University, Krijgslaan 281-S3, Gent 9000, Belgium}

\author{Emiliano Cort\'{e}s}
\affiliation{Chair in Hybrid Nanosystems, Nano-Institute Munich, Faculty of Physics, Ludwig-Maxilimians-Universit\"at M\"unchen, M\"unchen 80539, Germany}

\author{Stefan A. Maier}
\affiliation{Chair in Hybrid Nanosystems, Nano-Institute Munich, Faculty of Physics, Ludwig-Maxilimians-Universit\"at M\"unchen, M\"unchen 80539, Germany}

\author{Stefano Vezzoli}
\affiliation{The Blackett Laboratory, Department of Physics, Imperial College London, London SW7~2BW, United Kingdom}

\author{Iwan Moreels}
\affiliation{Department of Chemistry, Ghent University, Krijgslaan 281-S3, Gent 9000, Belgium}

\author{Riccardo Sapienza}
\email[To whom correspondence should be addressed; E-mail:]{ r.sapienza@imperial.ac.uk}
\affiliation{The Blackett Laboratory, Department of Physics, Imperial College London, London SW7~2BW, United Kingdom}

% Include the date command, but leave its argument blank.
%\date{}

\begin{abstract}
 {Electron transfer to an individual quantum dot promotes the formation of charged excitons with enhanced recombination pathways and reduced lifetimes. 
Excitons with only one or two extra charges have been observed and exploited for very efficient lasing or single quantum dot LEDs. 
Here, by room-temperature time-resolved experiments on individual giant-shell CdSe/CdS quantum dots, we show the electrochemical formation of highly charged excitons containing more than twelve electrons and one hole. 
We report the control over intensity blinking, along with a deterministic manipulation of quantum dot photodynamics, with an observed 210-fold increase of the decay rate, accompanied by 12-fold decrease of the emission intensity, while preserving single-photon emission characteristics. 
These results pave the way for deterministic control over the charge state, and room-temperature decay-rate engineering for colloidal quantum dot-based classical and quantum communication technologies.} 
% {Teaser: The rate of photon emission from a quantum dot can be controlled and speed-up by over hundred times by applying a voltage}
\end{abstract}

\maketitle

\noindent
\section{INTRODUCTION} 

The observation of reduced Auger recombination, leading to an increased quantum yield of colloidal quantum dots, has sparked a fast-paced progress in the development of highly fluorescent and stable quantum dots for displays~\cite{Steckel2015}, light-emitting diodes~\cite{Lim2018}, coherent~\cite{Kozlov2019} as well as quantum light sources~\cite{Lin2017,Pisanello2013}. 
Still, in the colloidal quantum dot community there is an ongoing struggle to reconcile suppressed Auger recombination with fast radiative recombination~\cite{Bae2013}. Indeed, existing systems that reduce Auger recombination also tend to have a lower electron-hole overlap~\cite{Bae2013}, and therefore  increased fluorescence lifetime (surpassing 100 ns~\cite{Dubertret2014,Christodoulou2014}), which hampers quantum and classical photonic technologies that rely on high brightness and fast communication rates.

A nanophotonic approach can boost light-matter interactions and modify an emitter's decay rate by several orders of magnitude~\cite{Koenderink2017,Nicoli2019,Mignuzzi2019}. Nevertheless, experimental studies have been limited so far to decay rate enhancements of $\sim 6$ for a quantum dot surrounded by a plasmonic shell~\cite{Dubertret2014}, or $\sim 80$ inside plasmonic nanogaps~\cite{Yuan2013} and patch antennas~\cite{Belacel2013}. This enhancement comes at the cost of a reduced single-photon emission purity due to strong  biexciton emission, limited tunability, and the fabrication challenge of nanometric precision in positioning the quantum dots~\cite{Morozov2018,Koenderink2017}.

A different route is to exploit exciton charging to enhance the emission rate of quantum dots themselves, which can be realized by electrochemical~\cite{Ding2002,Jha2010,Qin2011,Galland2011} or photochemical~\cite{Rinehart2013,Wu2017} charge injection.
 {Excitons with only one or two extra charges have allowed for the development of very efficient quantum dot lasing~\cite{Kozlov2019} and the understanding of blinking dynamics~\cite{Galland2011}, while charge transfer management has yielded single quantum dot LEDs~\cite{Lin2017}, LEDs with reduced efficiency roll-off~\cite{Lim2018}, and enabled studies of carrier and spin dynamics~\cite{Ferne2012}.}
The additional charge brings new recombination pathways ---  thus faster decay rates --- and modifies the electronic state of the quantum dot due to Coulomb interactions, which are enhanced by strong spatial confinement and reduced dielectric screening~\cite{Park2014}. 

Electrochemical injection of up to eight electrons in $1\mathrm{S}_e$ and $1\mathrm{P}_e$ states has been reported for thin ZnO \cite{Roest2002}, CdTe \cite{Jin2009}, PbSe \cite{Wehrenberg2005,Boehme2016}, and CdSe~\cite{GuyotSionnest2003,Boehme2013,Boehme2016,Spittel2017} quantum dot films. 
In these cases, charge injection in the lowest quantum state has been verified by a bleaching of the ground state exciton absorption.
Extension of such electrochemical charging experiments to individual quantum dots, beyond ensemble averaging, has been hampered by sample degradation at high voltages and poor photostability of the quantum dots. 
This has been remedied by exploiting the giant-shell quantum dot architecture, whereby  different emissive quantum states have been resolved in doubly charged CdSe/CdS and CdSeS/ZnS quantum dots, showing a reduced blinking and a modulated photoluminescence intensity and lifetime~\cite{Jha2010,Qin2011,Galland2011,Galland2012}. 

Here we go beyond the weak charge injection regime, and we report the observation of controllable, stable, highly-charged excitonic states in an individual giant-shell CdSe/CdS quantum dot.  {To induce the highly charged states, we used a lithography-free electrochemical cell.} We show reversible control of individual quantum dot single-photon dynamics, allowing for an on-demand increase of spontaneous emission decay rate up to 210-fold with only a minor 12-fold decrease in emission intensity. 

\section{RESULTS} 
\noindent
\textbf{Quantum dots and experimental setup} \\
Two batches of giant-shell CdSe/CdS quantum dots were synthesized following a recently published protocol \cite{Christodoulou2014} with a minor modification (see Methods). Both batches have the same 4~nm CdSe core, but a different shell thickness resulting in a total diameter of $10.6 \pm 1.1$~nm and $13.1 \pm 2.1$~nm, and we labeled them batch 1 and 2, respectively.  {Having two batches with different shell sizes aided us with the statistics, extension and reliability of our findings.}  A representative transmission electron microscope (TEM) image for batch 2 is shown in Fig.\ref{fig:1}a, together with the absorption and emission spectrum in Fig.\ref{fig:1}b showing the CdS band edge lying around 500~nm and an emission peak around 655~nm. These quantum dots exhibit non-blinking behavior at a low pump fluency (see Methods).

We excited individual quantum dots with a blue laser at 442~nm (2.8~eV) in a custom-built confocal microscope capable of recording the fluorescence with time correlated single photon counting  {with an overall time response of 400~ps}. The quantum dots were subjected to a voltage bias in an electrochemical cell, composed of a transparent ITO working electrode and Pt quasi-reference and counter electrodes (Fig.\ref{fig:1}c) as detailed in the Methods section. The position of the Fermi level (orange dotted line in Fig.\ref{fig:1}d) was controlled by the applied voltage bias. 
\begin{figure}[ht]
    \includegraphics[width=1\linewidth]{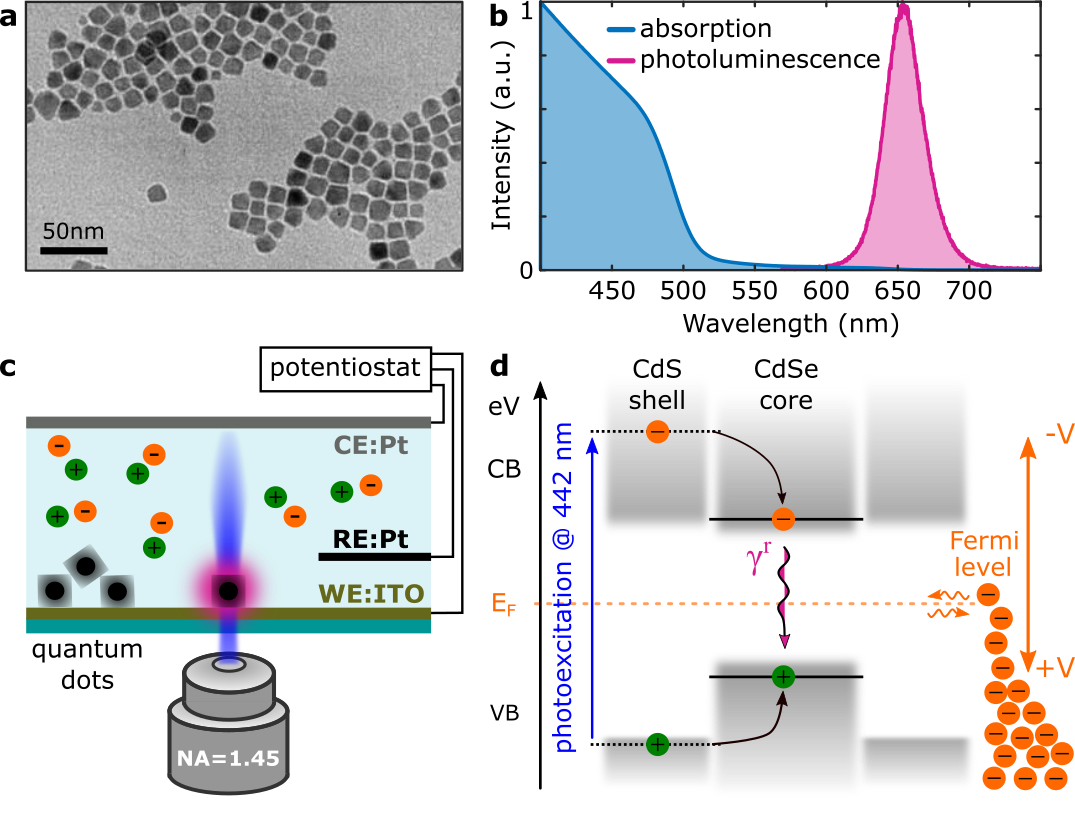}
	\caption{\textbf{Colloidal giant-shell CdSe/CdS quantum dots, and experimental setup.}
	\textbf{a} TEM image of batch 2.
	\textbf{b} Absorption and photoluminescence spectra of batch 2. 
	\textbf{c}~Sketch of a confocal microscope focused on an individual quantum dot subjected to a voltage bias between reference (RE) and working (WE) electrodes of a three-electrode electrochemical cell, while  a Pt coiled wire served as a counter  electrode (CE). 
	\textbf{d}~Energy diagram of a quantum dot  with valence (VB) and conduction (CB) bands accommodating an exciton. A 442 nm laser induces above bandgap excitation of the carriers in the CdS shell, which can relax to the CdSe core recombining radiatively at $\gamma^{r}=\gamma_{0}$ rate. The position of the Fermi level (orange dotted line) can be manipulated via the application of a voltage bias, and adjusted for the electron injection into the conduction band leading to exciton charging.
	}
	\label{fig:1}
\end{figure}

\noindent
\textbf{Statistical scaling model for charged excitons}\\
The optical properties of an individual quantum dot depend drastically on its charge state. The most common model describing the change in optical response under charging is the statistical scaling model~\cite{Galland2011,Sampat2015}. The model links the total decay rate ($\gamma_{N-1}$) and the quantum yield ($QY_{N-1}$) of  excitonic states with $N$ charges. In our case, because of the fast hole Auger rate in giant quantum dots, which is due to the stronger confinement~\cite{Vaxenburg2016}, we consider only the case of an excess of electrons. 
In this case, the radiative recombination rate of a charged exciton formed by the coupling of $N$ electrons in the conduction band and a single hole in the valence band increases with the electron number as $N \gamma_{0}$, where $\gamma_{0}$ is the  radiative  rate  of  a  neutral  exciton. It can be understood as an $N$-fold increase of the recombination pathways as each electron contributes. This is illustrated in Fig.\ref{fig:scamod}, where $\mathrm{X}_0$ (orange) is the neutral exciton, and $\mathrm{X}_-$ (green) is the negative trion.
Auger recombination is a non-radiative decay pathway, with a rate that increases  with the electron number  as $\gamma_{N-1}^A=N (N-1)\gamma^A$, where $\gamma^{A}$ is a constant characterizing the rate of a single electron pathway~\cite{Philbin2018}. The total recombination rate of band-edge excitons is the sum of radiative and non-radiative (Auger) rates, i.e. $\gamma_{N-1} = N\gamma_0 + \gamma_{N-1}^A$. Its statistical scaling can be rewritten as (see Supplementary Material for a full derivation):
\begin{eqnarray}
\gamma_{N-1}/\gamma_0=N[ 1+\frac{\gamma^A}{\gamma_0} (N-1) ],
\label{eq:gamscale}
\end{eqnarray}
and 
\begin{eqnarray}
QY_{N-1} = [1 +\frac{\gamma^A}{\gamma_0} (N-1) ]^{-1}.
\label{eq:qyscale}
\end{eqnarray}

\begin{figure*}[tbh]
	\includegraphics[width=1\linewidth]{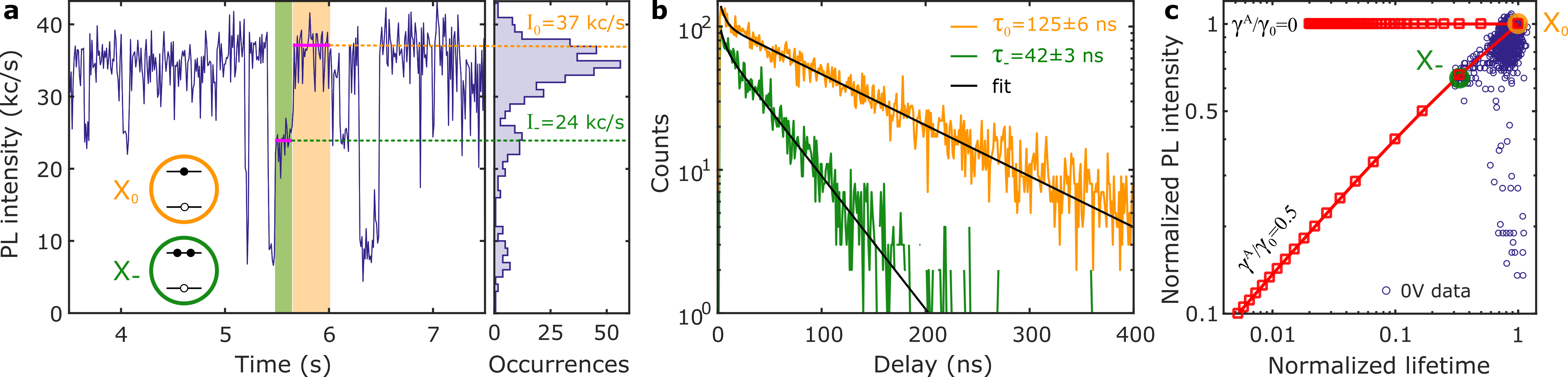}
	\caption{\textbf{Fluorescence lifetime intensity distribution~(FLID) at 0~V, and statistical scaling model.} 
	\textbf{a}~Photoluminescence intensity time trace of an individual quantum dot  {\textit{qd1} from batch 1} measured at 0~V bias and 30~nW excitation power. Neutral exciton  $\mathrm{X}_0$ and negative trion $\mathrm{X}_{-}$ are highlighted by orange and green areas, respectively.  {The time bin is 10~ms.} %
	\textbf{b}~Decay histograms of $\mathrm{X}_0$ and $\mathrm{X}_{-}$ were acquired using the photon arrival times from the orange and green time windows in \textbf{a}.
	The decay histograms are fitted with a bi-exponential function (black lines) in order to filter out the contribution of positive trion and multi-exciton states characterized by short lifetimes  (see Supplementary Material  {Fig.S5}). 
	\textbf{c} FLID visualizes $\mathrm{X}_0$ and $\mathrm{X}_{-}$ states with lifetime and intensity extracted from the selected green and orange time windows in \textbf{a}.
	The blue open circles represent statistics acquired during 60~s at 0~V bias.   {The red squares represent the discrete statistical scaling model (with a red connecting line as a guide to the eye), according to Eqs.~\ref{eq:gamscale} and \ref{eq:qyscale} calculated per additional electron, and its solutions are presented for the Auger-free case ($\gamma^A/\gamma_0 = 0$), and a quantum dot with  Auger processes ($\gamma^A/\gamma_0 = 0.5$)}.
	}
	\label{fig:scamod}
\end{figure*}

The Auger processes can be reduced in quantum dots with a giant shell \cite{Galland2012}. According to Eqs.~\ref{eq:gamscale} and \ref{eq:qyscale}, in the limit $\gamma^A \ll \gamma_0$, the emission rate of charged excitons roughly scales as $N\gamma_0$, and the emission intensity is similar to that of the neutral exciton $\mathrm{X}_0$. 

A typical experimental intensity time trace collected on a quantum dot of batch 1, for 0~V applied bias, is shown in Fig.\ref{fig:scamod}a (blue trace).  The decay histograms (Fig.\ref{fig:scamod}b) were extracted from Fig.\ref{fig:scamod}a by accumulating delay times in the two time windows indicated by the orange and green shaded areas. They reveal the neutral exciton $\mathrm{X}_0$ and the negative charged exciton (trion) $\mathrm{X}_{-}$ as the dominant states, with single-exponential decays and lifetime of $125 \pm 6$~ns and $42 \pm 3$~ns, respectively. These states can be identified in a fluorescence lifetime intensity distribution~(FLID), which correlates the fluorescence intensity and lifetime as shown in~Fig.\ref{fig:scamod}c. The blue points in the FLID were obtained by splitting the intensity time trace in 20~ms long time bins and computing the corresponding lifetime-intensity pair. The spread of the data points between $\mathrm{X}_0$ an $\mathrm{X}_-$ is due to the fast blinking (flickering), and thus averaging, between the two states. States observed with the same lifetime and different fluorescence intensity are an example of B-type blinking due to the hot-exciton trapping~\cite{Galland2012}. The $\mathrm{X}_0$ and $\mathrm{X}_{-}$ states are located along the red line, which connects the states predicted from Eq.~\ref{eq:gamscale} and Eq.~\ref{eq:qyscale}, for  {$\gamma^A/\gamma_0 = 0.5\pm0.1$}. The latter value is extracted from the $\mathrm{X}_0$/$\mathrm{X}_{-}$ lifetime-intensity ratios in Fig.\ref{fig:scamod}b after integrating for the orange and green time window periods. The relatively slow Auger recombination  {($\gamma^A/\gamma_0 = 0.5\pm0.1$)} is due to the thick 3.3~nm CdS shell in batch 1, and it is even slower  {($\gamma^A/\gamma_0 = 0.08\pm0.02$)}, thus less efficient, for quantum dots in batch 2 with 35\% thicker shell (4.5~nm, {\it{cfr.}} below).\\

\noindent
\textbf{Control of intensity blinking} \\
The intensity time trace at 0~V in Fig.\ref{fig:three}a presents a typical blinking behavior between high (35~kc/s) and low (5~kc/s) emissivity states, which correspond to the neutral exciton and positive trion (see Supplementary Material  {Fig.S5}). 
When the applied bias is lowered to $-1.4$~V (Fig.\ref{fig:three}c), the blinking between exciton and positive trion is completely suppressed, while the blinking between negatively charged excitons still takes place, which can be seen from the wide intensity distribution in the corresponding occurrences histogram. Further lowering of  the applied bias to $-1.7$~V induces formation of highly charged and stable excitonic states (Fig.\ref{fig:three}e), since the applied static bias does not allow them to decay to lower charged excitonic states.
Besides the reduction of blinking, the applied voltage has an effect on the quantum dot fluorescence intensity, as the average state emissivity decreases from 35~kc/s to 20~kc/s for $-1.4$~V. This dimming progresses further when the applied bias is lowered to $-1.7$~V, reaching 10~kc/s (Fig.\ref{fig:three}e). By Hanbury Brown and Twiss interferometry we verified that the second-order correlation at zero delay times $g^{(2)}(0)$ does not rise above 0.5 for the bias above -1.8~V, which means that the investigated quantum dot remains a single photon emitter (Fig.\ref{fig:three}b,d,f). Instead, at -1.8~V  $g^{(2)}(0)$ raises to $0.57\pm0.05$ (see Supplementary Material  {Fig.S4}).  The increase of the zero delay peak is assigned to an increased biexciton emission efficiency at high negative bias, as the Auger rate for holes becomes comparable to the radiative rate of charged excitons \cite{Manceau2014}.
For completeness, we report that some quantum dots displayed a photoluminescence brightening when subjected to a negative bias (see Supplementary Material  {Fig.S9}). 
We attribute this to a retrieval of the exciton brightness at a negative potential similar to what has been reported~in~\cite{Gooding2008}, probably due to an initial high density of defect states which were passivated. \\

\begin{figure*}[tbh]
    \includegraphics[width=0.95\linewidth]{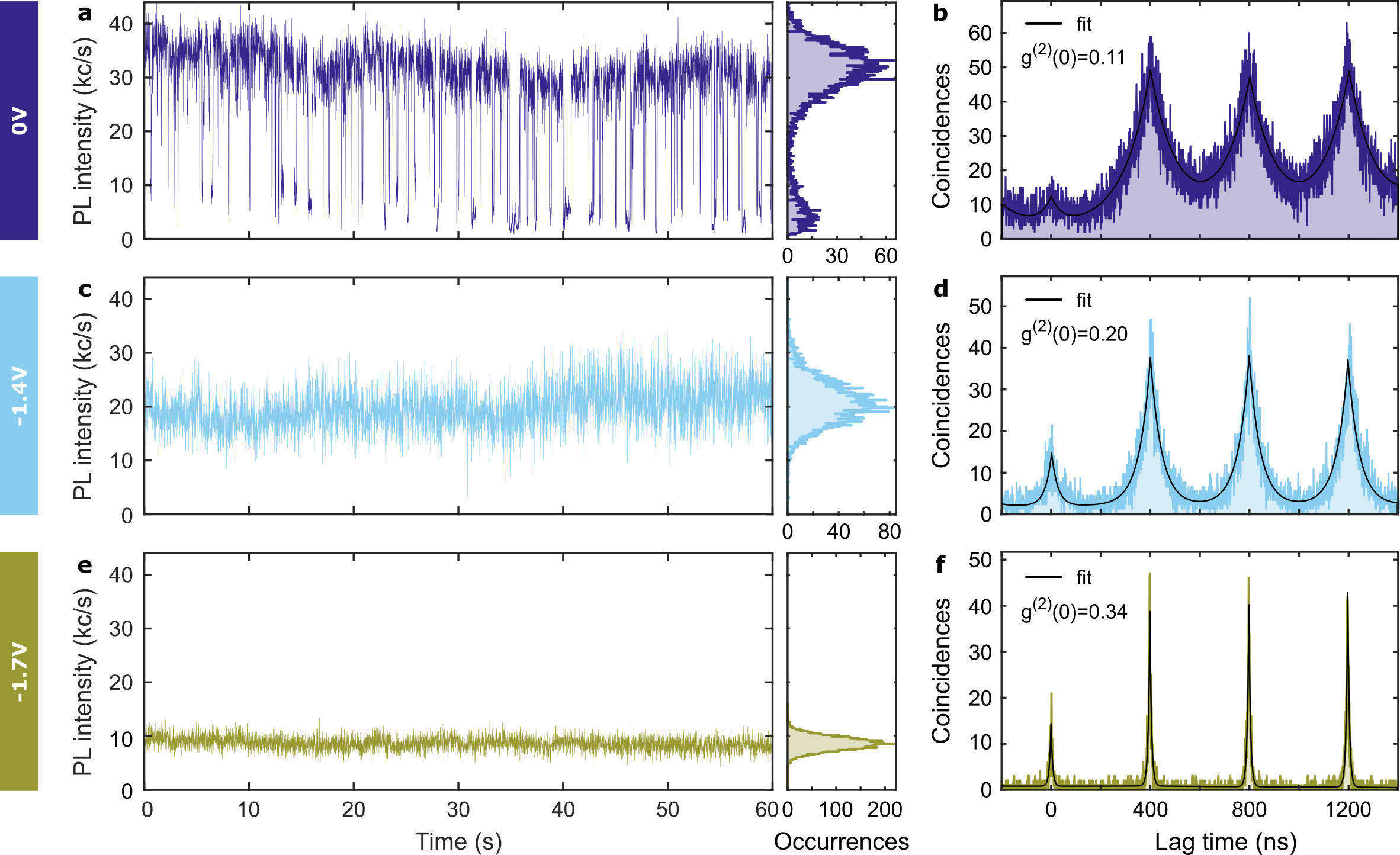}
	\caption{
	\textbf{Control of blinking and intensity via application of voltage bias. }
	\textbf{(a,c,e)} Photoluminescence intensity time traces of the quantum dot  {\textit{qd1} from batch 1} obtained at 0~V, -1.4~V, and -1.7~V static bias.  {The time bin is 10~ms.}
	\textbf{(b,d,f)} Intensity auto-correlation $g^{(2)}(t)$ histograms for the corresponding values of static voltage bias measured during 120~s. The measured values of $g^{(2)}(0)$ confirm that the quantum dot remains a good single photon source at negative bias. 
	\textbf{(a)}~At the constant bias of 0~V, we observe photoluminescence blinking between high and low emissive states, which we attribute to neutral exciton and positive trion.
	\textbf{(c)}~The decrease of applied potential to -1.4~V suppresses the blinking to positive trion. 
	\textbf{(e)}~Further lowering of bias to -1.7~V induces formation of highly charged excitons characterized by lower emission intensity. 
}
	\label{fig:three}
\end{figure*}

\begin{figure*}[tbh]
    \includegraphics[width=0.95\linewidth]{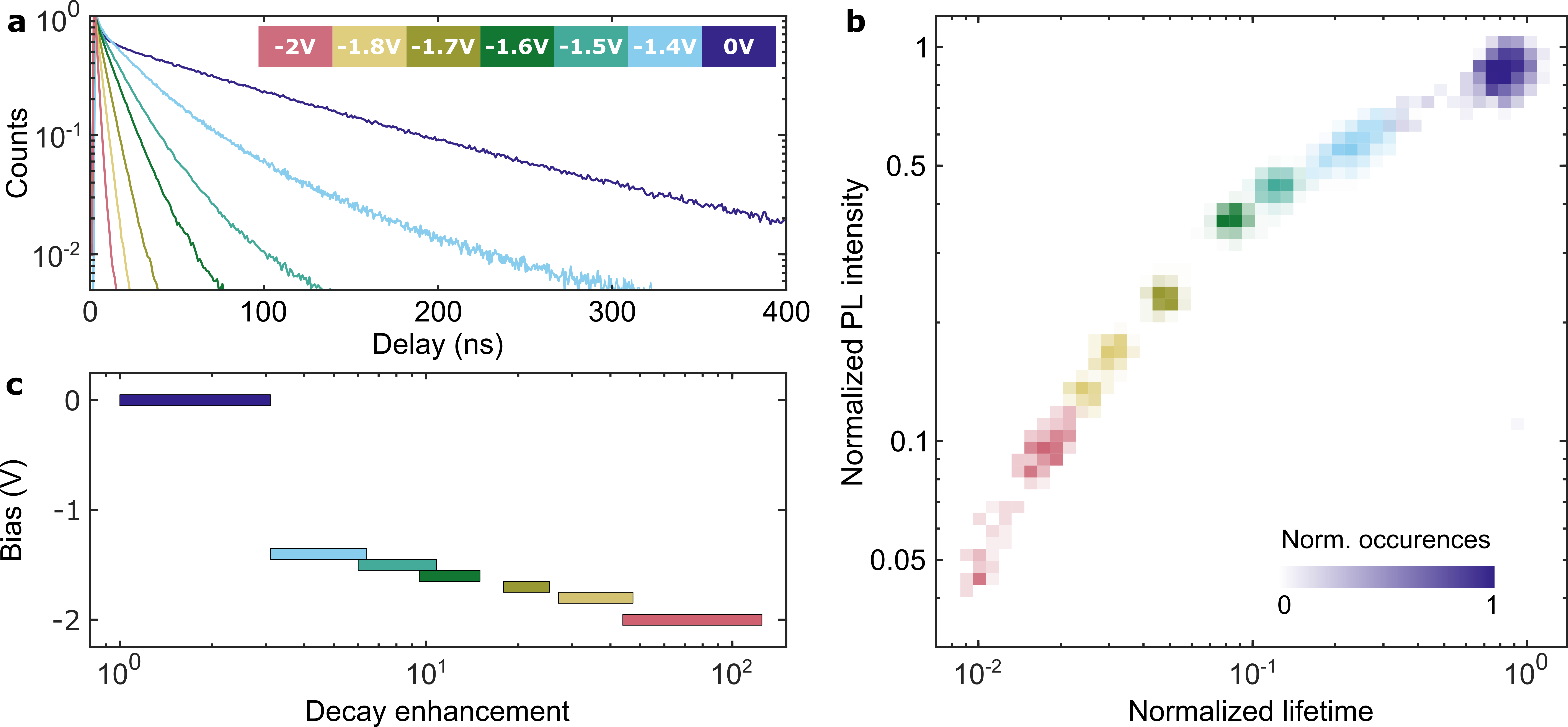}
	\caption{\textbf{Active control of decay rate with voltage bias.} 	
	 \textbf{a}~The quantum dot  \textit{qd1} from batch 1 was measured at different voltage bias in the electrochemical cell (the applied voltage bias are coded with colors as shown in the inset). The overall decay histograms acquired during 60~s demonstrate a shortening of the lifetime with increasing negative bias.
	 \textbf{b}~Each of the fluorescence intensity time traces, measured at static bias, was processed as described in Fig.\ref{fig:scamod} resulting in a FLID.  
	 The intensity-lifetime pairs are represented here as a normalized distribution where the number of occurrences is measured by the level of transparency (a representative scale bar is shown for 0~V).
	 \textbf{c}~Decay rate enhancement could be controlled by applying a voltage bias. The decay rate increased rapidly for voltages below -1.4~V. The shortest decay lifetime  $0.9\pm 0.2$~ns was measured at -2~V for this particular quantum dot, which corresponded to an enhancement of $140\pm 30$.
	}
	\label{fig:flid_of_V}
\end{figure*}
\begin{figure*}[tbh]
	\includegraphics[width=1\linewidth]{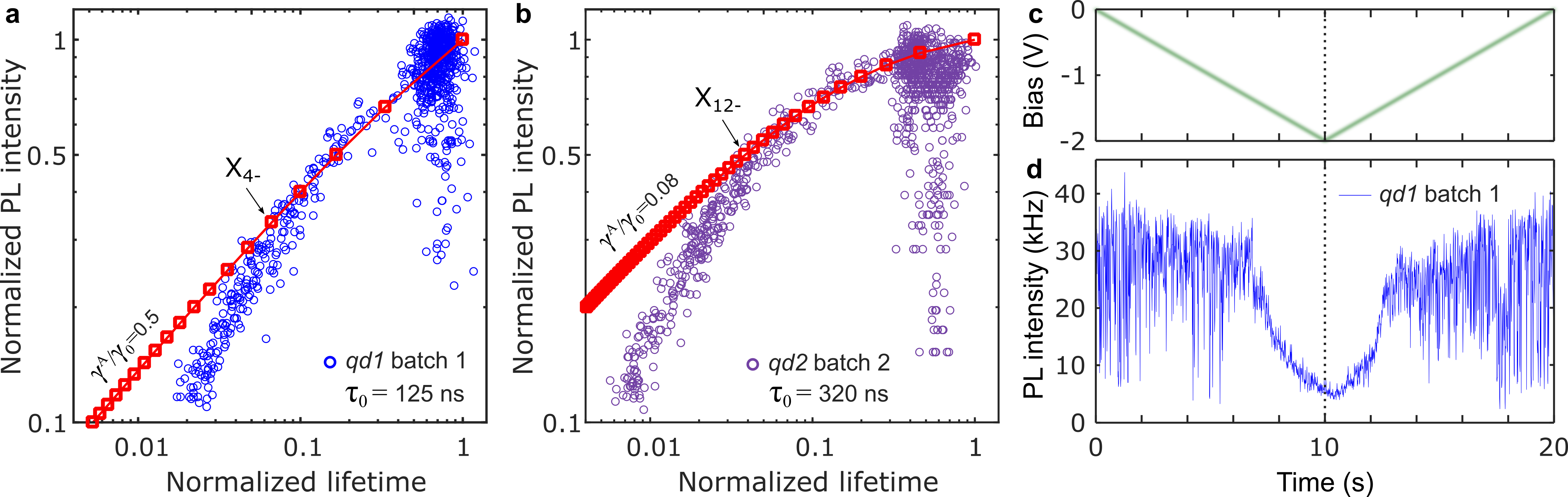}
	\caption{\textbf{Optical response of a quantum dot during a cyclic voltammetry scan demonstrating reproducibility.}
	\textbf{a}~FLID of the quantum dot (\textit{qd1}, blue circles) from batch 1, which has been presented in  {Fig.2-4}, extracted from intensity time traces measured during cyclic voltammetry scans as shown in \textbf{d}. \textbf{b} FLID of another quantum dot, from  batch 2 (\textit{qd2}, purple circles).  In both panels the red lines represent the statistical scaling model for different Auger rates  {according to Eqs.~\ref{eq:gamscale} and \ref{eq:qyscale}}. In the cyclic voltammetry scans the bias was varied linearly in time as indicated in \textbf{c}. At $t=10$~s the scan polarity was reversed to return to the initial bias of 0~V. Positive bias  {caused quenching of the neutral exciton (see an example in Supplementary Material Fig.S6}).
	 \textbf{d}~Intensity time trace measured during the potential scan of \textit{qd1} from batch 1 used to build FLID in panel~\textbf{a}. 
	 After the suppression of blinking around -1.4~V, the photoluminescence intensity was gradually quenched under the linear decrease of the applied potential. 
	 The initial photoluminescence intensity was restored when the scan polarity was reversed (an intensity trace of this quantum dot during 6 voltammetric cycles is shown in Supplementary Material  {Fig.S8}), and began to blink again around -1.4~V.  {The time bin is 10~ms.}
	 }
	\label{fig:mainres}
\end{figure*}

\noindent
\textbf{Control of decay rate}\\
The decrease of emission intensity shown in Fig.\ref{fig:three} is correlated with a change in decay rate. This is demonstrated in Fig.\ref{fig:flid_of_V}a for a wide range of negative voltage bias. Decay histograms integrated for 60~s for the same quantum dot as in Fig.\ref{fig:three} clearly indicate a drastic reduction of the fluorescence decay time, from 125~ns at $0$~V down to 0.9~ns at $-2$~V (140-fold). 
The correlation between fluorescence lifetime and intensity at various negative bias can again be combined into a FLID plot, as shown in Fig.\ref{fig:flid_of_V}b. The applied bias are encoded in different color maps that represent the distribution of occurrences. The measured states monotonically decrease their lifetime and intensity upon applying of the negative bias. For each applied voltage the quantum dot is in a well-defined lifetime-intensity state with fluctuations mostly due to experimental noise. \\

\noindent
\textbf{Charge state tomography of an individual quantum dot}\\
Fig.\ref{fig:mainres}a,b plot the FLID data for both batches, when the bias was continuously varied in a voltage scan from 0~V to -2~V as shown in Fig.\ref{fig:mainres}c. Cyclic voltammetry scan allowed us to span the charge state configuration space in a short time (20~s) minimizing optical misalignment issues. Obtained results were similar to what is observed at static bias in Fig.\ref{fig:flid_of_V}. 

The red line in Fig.\ref{fig:mainres}a,b represents the statistical scaling model according to Eq.~\ref{eq:gamscale}-\ref{eq:qyscale}, where the relative Auger rates $\gamma^A/\gamma_0$ were obtained, for batch 1, from the lifetime-intensity ratios of exciton and trion, as described in the section on Statistical scaling model for charged excitons. Instead, for batch 2 we extracted $\gamma^A/\gamma_0$ from the fit of the lifetime-intensity distribution of the neutral and low charged excitons, i.e. beyond the trion, as the quantum yields of exciton and trion are too similar and the Auger rate estimation has a very large error (see Supplementary Material Fig.S12). 
Remarkably, the evolution of intensity and lifetime for low bias, up to $\mathrm{X}_{4-}$ for batch 1 and $\mathrm{X}_{12-}$ for batch 2, can be fitted with the statistical scaling model, which is in essence a single-particle model that assumes a fixed electron-hole overlap and thus neglects many-body Coulomb interactions~\cite{Park2014}. For  lower negative bias, the quantum dots in batch 1 and 2 are charged beyond $\mathrm{X}_{4-}$ and $\mathrm{X}_{12-}$ states, respectively, and a pronounced deviation is clearly visible in Fig.\ref{fig:mainres}a,b  {(see Methods)}. In this case the FLID cannot be fitted with Eq.~\ref{eq:gamscale}-\ref{eq:qyscale}. Considering that previous results on charged excitons and biexcitons already confirmed that an ad-hoc $\beta$ factor should be used to improve the agreement between lifetime and fluorescence intensity in the scaling model \cite{Park2014,Sampat2015}, we postulate that the higher carrier densities created here will only lead to a further modification of the electron wave function, affecting both electron-hole overlap (and thus $\gamma_0$), as well as the Auger recombination rate $\gamma^A$, which is highly sensitive to the behavior of the electron wave function at the CdSe/CdS interface \cite{Climente2012}.
The lowest applied potential of -2~V causes a large decrease of the lifetime, with largest recorded values of $140\pm30$-fold, while the intensity drops by a factor of $25\pm4$ for batch 1, and $210\pm40$ and $12\pm3$ for batch 2, respectively. These values of decay enhancement and intensity drop were limited by the lowest voltage bias -2V, which did not cause the degradation of ITO substrate.

\noindent
\section{DISCUSSION}
{Coulomb repulsion upon charging, especially in air or vacuum, can limit the charge state attainable. However, in our experiments in liquid, adsorption of tetrabutylammonium (TBA) cations compensate the electrochemical build up surface potential due to the injected charges \cite{Puntambekar2016}. These screened charges reduce the overpotential requirements for further charge injection \cite{Boehme2013}. In a simplified estimation for our giant-shell quantum dots, each quantum dot (diameter $\sim 12$~nm) can allocate up to $\sim100$ TBA cations (radius $\sim 0.5$~nm) on its surface. Therefore, we conclude that Coulomb repulsion is not the dominant effect due to the larger surface area in our quantum dots}.

{Once the bias is increased to reach the band edge, electron injection depends on the available states. From a density of states reasoning, we have calculated the expected level spacing (details in Supplementary Material) and we confirm theoretically that for a voltage of 80-150~meV above the conduction band edge $\sim 20$ states can be populated.} Switching of the photodynamics from a charging state to another can be done by simply applying the voltage, with a rise time of about 2~$\mu$s, limited by our electronics (see Supplementary Material  {Fig.S8}). 

Finally, we discuss the repeatibility and reproducibility of the results.
We repeated this experiment with  {37} individual quantum dots  {from two batches}, and we observed charging beyond doubly negative charged exciton in  {13 cases, which is around} 30\%, while in the other cases the quantum dots did not demonstrate the lifetime-intensity dip as in Fig.5d (see Supplementary Material  {Fig.S7 and Tab.S1}).
Moreover, it has been recently pointed out that charging-induced damage can occur due to reduction of the quantum dot surface~\cite{fosse2019}.
Here, the application of up to -2~V of negative potential is reversible and does not damage the quantum dot (Fig.\ref{fig:mainres}d). We believe that this is because the quantum dot thick shell can accommodate many defect states. In our experiments the voltage bias was gradually varied, and Fig.\ref{fig:mainres}d shows a clear drop in the emitted intensity for voltages below {-1.4~V}  which recovers when the bias is returned to 0~V. This recovery can be repeated many times with no sign of degradation in the optical properties (Supplementary Material  {Fig.S8}): we tested it up to 540 cycles during 3 hours, a time-span which was limited by the degradation of the ITO working electrode.  {Besides}, {here the heating of  quantum dots is not a concern, as  we are not inducing photon absorption as in photo-excitation experiments.}

Controlling the charge state of an individual quantum dot can be very important for quantum technologies, where the undesired switching to a different charge state precludes interfacing the electronic spin to photons~\cite{Widmann2019}.
Boosting the decay rate now brings the colloidal nanocrystals on par with the fluorescence lifetime of NV-centers in diamond \cite{Mizuochi2012}, and epitaxial quantum dots \cite{Senellart2017}, and could open a path towards coherent emission at room temperature once the decay rate becomes faster than the decoherence rate~\cite{Accanto2016}.
An intensity-switchable nanoscale light source can find important applications for optical signal processing with very stable giant quantum dots, where the switching speed is usually limited by the decay rate of the quantum dot, which can here instead reach GHz speeds when the exciton is maximally charged.

In conclusion, we report the observation of highly charged excitons, which induces over 210-fold increase in the decay rate with only a 12-fold reduction of the quantum yield. The charging process is reversible and deterministic, allowing for direct manipulation of the quantum dot emission rate through the applied bias, while preserving the single photon emission characteristics.
 {The fluorescence intensity-lifetime relation observed at high charge density goes beyond the conventional statistical scaling model, and not all quantum dots show the lifetime reduction, indicating the need for a more general model including many-body corrections.}  
Charging colloidal quantum dots is a powerful route for enhancing and controlling their photodynamics, which has important implications for tunable quantum sources, brighter displays, optical signal processing and can lead to novel approaches to charge/voltage sensing.\\

\noindent
\section*{MATERIALS AND METHODS}
\noindent
\textbf{Quantum dots}\\
CdSe/CdS core/shell quantum dots were synthesized using established methods \cite{Christodoulou2014}, with a small modification: before CdS shell growth, CdSe core quantum dots were suspended in ODE together with 0.25~mL of a 0.5~M solution of cadmium oleate in ODE. This mixture was degassed for 30 min at 110$^\circ$C, and subsequently heated to 300$^\circ$C. Next, an equimolar mixture a 0.5~M cadmium oleate solution in ODE and a 0.5~M TOPS solution was then slowly injected (at a rate of about 1~mL per hour) by syringe pump to grow the CdS shell, with the total amount of Cd- and S-precursors determined by the desired shell thickness. 
{Individual quantum dots from both batches exhibited non-blinking photodynamics at low pump fluency, and  blinking at high pump fluency as in the presented here experiments (see Supplementary Material Fig.S2).}
 {We remark here that we used a very short time bin (10~ms) to plot the intensity time traces, much shorter than in most reported non-blinking quantum dots experiments (30-200~ms) \cite{Galland2012,Park2014,Meng2016,Hou2019}, which is also why we can capture very fast blinking events.}\\

\noindent
\textbf{Electrochemistry}\\
The experimental setup consisted of a custom-built three-electrode electrochemical cell mounted on a time-resolved confocal fluorescence microscope. Diluted quantum dots in toluene were spin-coated at ITO substrates (70-100 $\Omega / \Box$, Diamond Coatings), which was the working electrode of the electrochemical cell. Coiled and straight Pt wires served as counter and quasi-reference electrodes, respectively.  {The distance between working and counter electrodes was 0.5~cm.} The electrolyte was 0.1~M tetrabutylammonium hexafluorophosphate (TBAPF$_6$, Sigma-Aldrich, $\geq$99.0\%) in acetonitrile (Sigma-Aldrich, 99.8\%) or propylene carbonate {(Sigma-Aldrich, 99\%)}. The voltage bias between the reference and working electrodes was controlled with a CHI 760C potentiostat (CH Instruments). As an almost negligible Ohmic drop was determined for our setup (up to 9~mV, cf. notes in Supplementary Material), all voltage biases are reported as recorded, i.e. without $iR$ drop correction and versus the Pt quasi reference electrode. The Pt quasi reference electrode potential against Ag/AgCl(sat) was measured to be $57\pm2$~mV under our experimental conditions (cf. notes in Supplementary Material for conversion of the potential values into NHE scale).\\

\noindent
\textbf{Lifetime measurements}\\
We used a blue laser (LDH-D-C-440 PicoQuant) at 442~nm with a pulse width of 64~ps and a repetition frequency of 2.5~MHz to excite quantum dots. Samples were scanned using a three-dimensional piezo stage (E-545.3CD~PI~Nano). A high NA oil-immersion objective (Plan Apochromat 100x, NA=1.45) focused the laser beam on an individual quantum dot and collected the photoluminescence signal using an avalanche photo diode  (SPCM-AQRH, PerkinElmer) connected to a time-correlated single photon counting module (TimeHarp 260, PicoQuant).  Photoluminescence decay histograms were obtained by recording the time between a laser excitation pulse and arrival time of a photon at a detector. \\

\noindent
\textbf{Photon antibunching}\\
The collected photoluminescence signal from a quantum dot was tested in an Hanbury Brown and Twiss interferometer to verify the single photon emission nature. The setup consisted of a 50/50 beam splitter and two avalanche photo diodes, which detected the arrival times of photons to build a coincidence histogram.
The second-order correlation function $g^{(2)}(0)$ was measured by comparing the peak area at zero arrival time with the area averaged over the first 3 lagging peaks  {without any background subtraction}. \\

\noindent
 {\textbf{Charge state of exciton}\\
We extract the excitonic charge state by fitting the intensity-lifetime evolution with the statistical scaling model, which is valid for low charge states assuming a fixed electron-hole overlap and neglecting many-body effects. The fit is robust, we have tested it by varying the initial conditions and the fitting range and we got a variability in the range 4 to 6 electrons for batch 1 and 12 to 16 for batch 2 (see Supporting Material Fig.S12). We report a conservative estimation of the exciton charging (4 and 12 for batch 1 and 2, respectively), which was extracted from the point where the statistical scaling model starts to deviate strongly from the experimental results, which is $\mathrm{X}_{4-}$ for batch 1 and $\mathrm{X}_{12-}$ for batch 2 (see Fig.5a-b).}

\noindent
\textbf{Acknowledgments: }
S.M. and R.S. acknowledge funding by EPSRC (EP/P033369 and EP/M013812/1). 
E.C. and S.A.M. acknowledge funding and support from the Deutsche Forschungsgemeinschaft (DFG, German Research Foundation) under Germany's Excellence Strategy – EXC 2089/1 – 390776260, the Bavarian program Solar Energies Go Hybrid (SolTech) and the Center for NanoScience (CeNS).  {E.C. acknowledges funding from 2020 European Research Council 802989 CATALIGHT).} S.A.M acknowledges the Royal Society, AFOSR/EOARD, and the Lee-Lucas Chair in Physics. This project has received funding from the European Research Council (ERC) under the European Union’s Horizon 2020 research and innovation program (802989 CATALIGHT, 714876 PHOCONA). A.H.K. and I.M. acknowledge the TEM facility of the Nematology Research Unit, member of the UGent TEM-Expertise center (life sciences). \textbf{Author contributions: } R.S. conceived the idea, I.M., A.H.K. and A.P. synthesized the quantum dots, S.M., E.L.P. and E.C. designed the electrochemical cell, S.M. conducted the experiments, S.M., S.V. and R.S. analyzed the data. The project was supervised by R.S., I.M. and S.A.M. All authors provided critical feedback and helped shape the research, analysis and manuscript. \textbf{Competing interests:} The authors declare that they have no competing interests. \textbf{Data and materials availability: } All data needed to evaluate the conclusions in the paper are present in the paper and/or the Supplementary Materials.  {The data that support the findings of this study are openly available in Figshare at http://doi.org/10.6084/m9.figshare.12145593.}

\end{document}